\documentclass[12pt]{article}
\usepackage{amsmath}
\usepackage{amssymb}
\usepackage{color}

\def\hybrid{\topmargin 0pt	\oddsidemargin 0pt 
	\headheight 0pt	\headsep 0pt
	\textheight 9in		
	\textwidth 6.1in	
	\marginparwidth .875in
	\parskip 5pt plus 1pt	\jot = 1.5ex}

\hybrid

\makeatletter
\def\titlepage{\@restonecolfalse\if@twocolumn\@restonecoltrue\onecolumn
     \else \newpage \fi \thispagestyle{empty}\c@page\z@	
	\def\thefootnote{\fnsymbol{footnote}} }

\def\endtitlepage{\if@restonecol\twocolumn \else \newpage \fi
	\def\thefootnote{\arabic{footnote}} 
	\setcounter{footnote}{0}}  

\def\numberbysection{\@addtoreset{equation}{section}
	\def\theequation{\thesection.\arabic{equation}}}
\makeatother
\relax

\numberbysection

\DeclareMathOperator{\U}{U}
\DeclareMathOperator{\Orth}{O}
\DeclareMathOperator{\GL}{GL}
\DeclareMathOperator{\SOrth}{SO}
\DeclareMathOperator{\SL}{SL}
\DeclareMathOperator{\Spx}{Sp}

\DeclareMathOperator{\Ad}{Ad}
\DeclareMathOperator{\Isom}{Isom}
\DeclareMathOperator{\ad}{ad}
\DeclareMathOperator{\coframe}{\mathcal{F}}
\DeclareMathOperator{\Orthframe}{SO}
\newcommand{\Btil}{\widetilde{B}}
\newcommand{\Gtil}{\widetilde{G}}
\newcommand{\Htil}{\widetilde{H}}
\newcommand{\Mtil}{\widetilde{M}}

\newcommand{\Ptil}{\widetilde{P}}
\newcommand{\Rtil}{\widetilde{R}}
\newcommand{\Stil}{\widetilde{S}}
\newcommand{\Xtil}{\widetilde{X}}
\newcommand{\Jtil}{\tilde{J}}
\newcommand{\Vtil}{\widetilde{V}}

\newcommand{\bbR}{\mathbb{R}}

\newcommand{\dminus}{\partial_{-}}
\newcommand{\dplus}{\partial_{+}}
\newcommand{\xitil}{\tilde{\xi}}
\newcommand{\ftil}{\tilde{f}}
\newcommand{\gtil}{\tilde{g}}
\newcommand{\htil}{\tilde{h}}
\newcommand{\half}{\frac{1}{2}}
\newcommand{\hodge}{\ast_{\Sigma}}

\newcommand{\lieg}{\mathfrak{g}}
\newcommand{\liegtil}{\tilde{\mathfrak{g}}}

\newcommand{\nablatil}{\widetilde{\nabla}}
\newcommand{\omegatil}{\tilde{\omega}}
\newcommand{\omegahat}{\hat{\omega}}

\newcommand{\xtil}{\tilde{x}}

\begin{document}

\begin{titlepage}
\mbox{April 2002}\hfill\mbox{UMTG--236}\newline
\strut\hfill\mbox{\tt hep-th/0204011}
\par\vspace{.2in}
\begin{center}
    {\bf\large Pseudoduality in Sigma Models\footnote{This work
    was supported in part by National Science Foundation grants
    PHY--9870101 and PHY--0098088.}}\\[.2in]
    {\bf Orlando Alvarez\footnote{email: 
    \href{mailto:oalvarez@miami.edu?subject=Pseudoduality paper}{\texttt{oalvarez@miami.edu}}}}\\
    {\em Department of Physics}\\
    {\em University of Miami}\\
    {\em P.O. Box 248046}\\
    {\em Coral Gables, FL 33124 USA}\\
\end{center}
\begin{abstract}
We revisit classical ``on shell'' duality, \emph{i.e.}, pseudoduality,
in two dimensional conformally invariant classical sigma models and
find some new interesting results.  We show that any two sigma models
that are ``on shell'' duals have opposite $1$-loop renormalization
group beta functions because of the integrability conditions for the
pseudoduality transformation.  A new result states for any two compact
Lie groups of the same dimension there is a natural pseudoduality
transformation that maps classical solutions of the WZW model on the
first group into solutions of the WZW model on the second group.  This
transformation preserves the stress-energy tensor.  The two groups can
be non-isomorphic such as $B_{l}$ and $C_{l}$ in the Cartan notation. 
This transformation can be used for a new construction of non-local
conserved currents.  The new non-local currents on $G$ depend on the
choice of dual group $\Gtil$.

\end{abstract}
\setcounter{page}{1}
\end{titlepage}

\section{Introduction}
\label{sec:intro}

In this article we generalize the discussion in \cite{Alvarez:2000pk}
of classical ``on shell'' duality, also called
pseudoduality\footnote{This term was introduced in
\cite{Curtright:1994be} to distinguish from true ``off shell'' duality
where the duality transformation is canonical \cite{Giveon:1989tt,
Curtright:1994be, Alvarez:1994wj}.}, to the case where the nonlinear
sigma model has ``torsion'', see \emph{e.g.}, \cite{Nappi:1980ig,
Gates:1984nk,Curtright:1984dz,Braaten:1985is}.  An early example is
the pseudoduality between the non-linear sigma model on a group and
the pseudochiral model discovered by Zakharov and Mikhailov
\cite{Zakharov:1978pp}.  For notational conventions and for a more
complete set of references on duality, especially ``off shell''
duality inspired by string theory, see~\cite{Alvarez:2000pk}.

We take spacetime $\Sigma$ to be two dimensional Minkowski space.  The
sigma model with target space $M$, metric $g$ and $2$-form $B$ will be
denoted by $(M,g,B)$ and has lagrangian
\begin{equation}
    \mathcal{L} = \half g_{ij}(x)
    \left(
    \frac{\partial x^{i}}{\partial \tau}
    \frac{\partial x^{j}}{\partial \tau}
    -\frac{\partial x^{i}}{\partial \sigma}
    \frac{\partial x^{j}}{\partial \sigma}
    \right) +
    B_{ij}(x)
    \frac{\partial x^{i}}{\partial \tau}
    \frac{\partial x^{j}}{\partial \sigma}\,,
    \label{eq:lag}
\end{equation}
where $x:\Sigma \to M$ and the closed $3$-form $H$ is defined by
$H=dB$.  This theory is classically conformally invariant.  Our goal
is to see if we can relate solutions of the equations of motion of a
sigma model $(M,g,B)$ to the solutions of the equations of motions of
a different sigma model $(\Mtil,\gtil,\Btil)$.

Our default scenario is general riemannian manifolds but we often
specialize to the case of Lie groups.  Overall, the methods we use are
differential geometric ones that expand on ideas in
\cite{Alvarez:2000pk,Alvarez:2000bh,Alvarez:2000bi}.  The bundle of
orthonormal frames, the Cartan structural equations and the exterior
differential calculus play a central role.  Early work on using
differential form methods to study sigma models may be found in
\cite{DAuria:1980cx,DAuria:1980tb}.  In Section~\ref{sec:RG} we show
that a consequence of the integrability conditions for the existence
of the pseudoduality transformation is that any two sigma models that
are classically pseudodual have opposite $1$-loop renormalization group
beta functions.

Some of the most interesting explicit results involve specializing to
Lie groups and especially the classical ``strict'' WZW model
\cite{Witten:1984ar}.  This is the model with the Wess-Zumino term
normalized so that equations of motion are $\partial_{-}( g^{-1}
\partial_{+}g)=0$.  Given any two compact Lie groups of the same
dimension, we show that there is a duality transformation that maps
solutions of the equations of motion of the first strict WZW model
into solutions of the equations of motions of the second strict WZW
model.  The exposition of these specific results in
Sections~\ref{sec:example-abelian}, \ref{sec:example-nonabelian},
\ref{sec:WZWgeometry} and \ref{sec:conservation} is self-contained and
requires very little from the rest of the paper.

We revisit some ideas of Braaten, Curtright and Zachos
\cite{Braaten:1985is} on the geometry of sigma models and amplify and
clarify some issues in Section~\ref{sec:detour}.  We also revisit and
generalize some ideas presented by Ivanov \cite{Ivanov:1987yv} on
duality in sigma models with target spaces that are related to Lie
groups where he presents two lines of investigation.  The first has to
do with what could be called Pohlmeyer type duality
\cite{Pohlmeyer:1975nb} which is mostly tangential to our discussion. 
In the Pohlmeyer type duality, the ``duality'' equations are
schematically of the type $\partial_{\pm}\xtil =
e^{\pm\lambda}\partial_{\pm}x$ where $\lambda$ is a parameter.  A
systematic study of these relations leads, for example, to an infinite
number of conservations laws
\cite{Pohlmeyer:1975nb,Eichenherr:1979ci,Eichenherr:1981sk}.  Here we
adapt this construction to our case by observing that an initial
condition in the solution of an ordinary differential equation plays a
role similar to $\lambda$ and we use this to generate an infinite
number of conservation laws in Section~\ref{sec:conservation}.  The
second line of investigation deals with pseudoduality where the
pseudoduality equations are schematically of the form
$\partial_{\pm}\xtil = \pm \partial_{\pm}x$.  Here, we are interested
in this second type of duality.  
Ivanov studied sigma models associated with Lie groups but his 
formalism only allowed dual models with $\Htil=0$.
Our generalization of Ivanov's method to general riemannian manifolds
in Section~\ref{sec:geometry} will explain clearly why he could
only discuss the case $\Htil =0$ and also makes connection to results 
in \cite{Braaten:1985is}.

This article is organized as follows.  The basic framework is
established in Sections \ref{sec:framework} and
\ref{sec:pseudoduality}.  The main result of this paper is
eq.~\eqref{eq:connection} that relates the metrics and $3$-forms on
the respective manifolds.  The integrability conditions for
pseudoduality are discussed in Section~\ref{sec:integrability} along
with the connection to the renormalization group.  A variety of
explicit examples are discussed in Sections \ref{sec:examples} and
\ref{sec:strictWZW}.  Section~\ref{sec:geometry} studies the
differential geometry of some naturally occurring connections.  The
Appendices provide some background material.

\section{The Framework}
\label{sec:framework}

The formulation of the general duality transformation is best done in
the bundle of orthonormal coframes.  For a brief review of
$G$-structures see Appendix~\ref{sec:g-structures}.  The reader may
want to look in \cite{BCG3} and study their discussion about
isometries between Riemannian manifolds and try to understand the idea
behind E.~Cartan's technique of the graph~\cite{Warner}.  We first
discuss the problem locally and see how it becomes simpler and more
natural in the bundle of orthonormal coframes.  We begin with local
discussion of pseudoduality on $M$ and $\Mtil$ and then show how to
lift these concepts to the orthonormal coframe bundles $\Orthframe(M)$
and $\Orthframe(\Mtil)$.  A more mathematically rigorous discussion
would entail a discussion of jet bundles that we prefer to avoid.

Let $V$ and $\Vtil$ be local neighborhoods
respectively in $M$ and $\Mtil$.  In these neighborhoods choose local
orthonormal coframes $\{\omega^{i}_{V}\}$ and
$\{\omegatil^{i}_{\Vtil}\}$. The $\sigma^{\pm}$ derivatives of the 
sigma model maps $x:\Sigma \to M$ and $\xtil:\Sigma\to \Mtil$ are 
given by
\begin{equation}
    \omega^{i}_{V} = (x_{V})^{i}{}_{a}d\sigma^{a}
    \quad\mbox{and}\quad
    \omegatil^{i}_{\Vtil} = (\xtil_{\Vtil})^{i}{}_{a}d\sigma^{a}\,.
    \label{eq:deflocalvel}
\end{equation}
The pseudoduality equations  \cite{Alvarez:2000pk} are
\begin{equation}
    (\xtil_{\Vtil})_{\pm}(\sigma) = \pm T_{\pm}(\sigma)
    (x_{V})_{\pm}(\sigma)\,,
    \label{eq:psd1}
\end{equation}
where the matrices $T_{\pm}(\sigma)$ are in $\SOrth(n)$.  In this
article we only treat the case $T_{+}=T_{-}$.   Over the
neighborhood $V\subset M$ the bundle of coframes $\Orthframe(M)$ is
locally $V \times \SOrth(n)$.  A point may be given coordinates
$(x,R_{V})$ where $R_{V}\in \SOrth(n)$ is the matrix that describes
the coframe $\omega_{V}$ relative to a fiducial coframe.  We saw in
Appendix~\ref{sec:g-structures} that $\omega = R_{V}\omega_{V}$ is the
canonical $1$-form on $\Orthframe(M)$ and it is globally defined.  The
coframe bundle $\Orthframe(M)$ has a global coframing given by the
canonical $1$-forms $\omega^{i}$ and by the globally defined torsion
free riemannian connection $1$-forms $\omega_{ij}$, $\omega_{ij}=
-\omega_{ji}$.  These satisfy the Cartan structural equations
\begin{eqnarray}
    d\omega^{i} & = & -\omega_{ij}\wedge\omega^{j}\,,
    \label{eq:cartan1}  \\
    d\omega_{ij} & = & -\omega_{ik}\wedge\omega_{kj}
       + \half R_{ijkl} \omega^{k}\wedge\omega^{l}\,,
    \label{eq:cartan2}
\end{eqnarray}
where $R_{ijkl}$ are the Riemann curvature functions on the 
orthonormal coframe bundle\footnote{The 
Riemann curvature tensor on $M$ is equivalent to the globally defined 
curvature functions on $\Orthframe(M)$. In general, tensors on the 
base become functions on the coframe bundle.}. We emphasize that the 
set $\{\omega^{i},\omega_{jk}\}$ gives a global coframing of the 
coframe bundle $\Orthframe(M)$. Lastly we point out that if 
$(\omega_{V})_{ij}$ is the expression for the riemannian connection 
in a local coframe $\omega_{V}$ in $V\subset M$ then the globally 
defined $\omega_{ij}$ on $\Orthframe(M)$ is locally given by
\begin{equation}
    \omega_{ij} = (R_{V})_{ik}(\omega_{V})_{kl}(R_{V})^{-1}_{lj}
    - (dR_{V})_{ik} (R_{V})^{-1}_{kj}\;.
    \label{eq:globalconn}
\end{equation}
We also remind the reader that the local connection coefficients 
$(\omega_{V})_{ijk}$ are given by
\begin{equation}
    (\omega_{V})_{ij} = (\omega_{V})_{ijk} \omega^{k}_{V}\,.
    \label{eq:conncoeff}
\end{equation}
Up in the coframe bundle, $\omega^{i}$ and $\omega_{jk}$ are linearly 
independent and there is no relation analogous to (\ref{eq:conncoeff}).

If we look at (\ref{eq:deflocalvel}) we immediately see that 
$x^{i}{}_{a} = (R_{V})^{i}{}_{j}(x_{V})^{i}{}_{a}$ are globally 
defined functions on the coframe bundle $\Orthframe(M)$. Likewise we 
do a similar construction in $\Mtil$. In fact we 
see that
\begin{equation}
    \omega^{i} = x^{i}{}_{a}d\sigma^{a}
    \quad\mbox{and}\quad
    \omegatil^{i} = \xtil^{i}{}_{a}d\sigma^{a}\,.
    \label{eq:defvel}
\end{equation}
These are globally defined equations on the bundles of orthonormal 
coframes.

Let us do a warm-up first by describing the isometry problem in this
framework.  We are interested in finding an orientation preserving
isometry between $M$ and $\Mtil$.  We know that locally we need the
existence of a special orthogonal matrix valued function
$T:V\to\SOrth(n)$ such that $\omegatil_{\Vtil} = T_{V}\omega_{V}$. 
The isometry problem is formulated ``upstairs'' by asking whether we
can solve the pfaffian system of equations $\omegatil^{i}=\omega^{i}$ on
$\Orthframe(M)\times\Orthframe(\Mtil)$.  The reason it that locally
these equations  may be written as $\Rtil_{\Vtil}\omegatil_{\Vtil}=
R_{V}\omega_{V}$ and we see that a judicious choice of coframes will
give $T_{V} = (\Rtil_{\Vtil})^{-1}R_{V}$.

\section{The Pseudoduality Condition}
\label{sec:pseudoduality}

In this article we discuss the special case of equations
(\ref{eq:psd1}) where $T_{+}=T_{-}$.  Instead of thinking of $x:\Sigma
\to M$ you should think of a lift $X:\Sigma \to \Orthframe(M)$.  Thus
we have the pullbacks
\begin{equation}
    X^{*}\omega^{i} = x^{i}{}_{a}d\sigma^{a}
    \quad\mbox{and}\quad
    X^{*}\omega_{ij} = \omega_{ija}d\sigma^{a}
    \label{eq:derx}
\end{equation}
that define the derivatives.  From now on following the convention
used in exterior differential systems \cite{BCG3} we assume the
pullback is implicit, \emph{e.g.}, $\omega^{i} =
x^{i}{}_{a}d\sigma^{a}$.  Note that on $\Orthframe(M)$ the $1$-forms
$\omega^{i}$ and $\omega_{jk}$ are linearly independent so there is no
relation such as $\omega_{jk} = \Gamma_{ijk}\omega^{i}$ on
$\Orthframe(M)$.  If you use a (local) section $s:M\to \Orthframe(M)$
to pullback $\omega^{i},\omega_{jk}$ to $M$ then you would find
$s^{*}\omega_{jk} = \Gamma_{ijk} s^{*}\omega^{i}$ where 
$\Gamma_{ijk}$ are functions on $M$.  Define the second
derivatives of $x^{i}$ by
\begin{equation}
    dx^{i}{}_{a} + \omega_{ij}x^{j}{}_{a} = x^{i}{}_{ab}d\sigma^{b}\,.
    \label{eq:covder}
\end{equation}
By taking the exterior derivative of the first of (\ref{eq:derx}) you
learn that $x^{i}{}_{ab}= x^{i}{}_{ba}$.  Locally on $V\subset M$ we
have the $2$-form $B_{V}$ in the action (\ref{eq:lag}) with $H_{V} =
dB_{V}$.  The $3$-form is lifted to $\Orthframe(M)$ where it defines
functions $H_{ijk}$ such that $H = \frac{1}{3!}H_{ijk}
\omega^{i}\wedge\omega^{j}\wedge\omega^{k}$.  The equations of motions
may be written on the bundle as
\begin{equation}
    x^{k}_{+-} = - \half H_{kij} x^{i}{}_{+} x^{j}{}_{-}\,.
    \label{eq:eom}
\end{equation}
The stress energy tensor for the sigma model $(M,g,B)$ is given by
\begin{equation}
    \Theta_{+-}=0\,,\quad
    \Theta_{++}= x^{i}_{+}x^{i}_{+} \quad\mbox{and}\quad
    \Theta_{--}= x^{i}_{-} x^{i}_{-}\,.
    \label{eq:Tmunu}
\end{equation}
Of course there are similar equations on $\Orthframe(\Mtil)$.

Analogous to the isometry problem, the
pseudoduality equations on the bundle of orthonormal coframes become
\begin{equation}
    \xtil^{i}{}_{\pm} = \pm x^{i}{}_{\pm}\,.
    \label{eq:special}
\end{equation}
An important feature of these pseudoduality equations is that they
preserve the stress-energy tensor.  Taking the exterior derivative of
the above and using (\ref{eq:covder}) we see that
$$
    -\omegatil \xtil_{\pm} + \xtil_{\pm a} d\sigma^{a} =
    \mp \omega x_{\pm} \pm x_{\pm a}d\sigma^{a}\,.
$$
If we use the duality equations (\ref{eq:special}) we have
$$
    \mp \omegatil  x_{\pm} + \xtil_{\pm a}d\sigma^{a}
    =  \mp  \omega x_{\pm}
    \pm  x_{\pm a} d\sigma^{a}\,.
$$
A little algebra shows that
$$
     \xtil_{\pm a}d\sigma^{a}
    = \pm (-   \omega  + \omegatil )x_{\pm}
    \pm  x_{\pm a} d\sigma^{a}\,.   
$$
We wish to isolate the integrability conditions so wedge the above 
with $d\sigma^{\pm}$.
$$
     \xtil_{\pm \mp}d\sigma^{\mp} \wedge d\sigma^{\pm}
    = \pm ( -   \omega  + \omegatil )x_{\pm} \wedge d\sigma^{\pm}
    \pm  x_{\pm \mp} d\sigma^{\mp} \wedge d\sigma^{\pm}\,.   
$$
We have two equations
\begin{eqnarray*}
     \xtil_{+ -}d\sigma^{-} \wedge d\sigma^{+}
    &=& + (  -   \omega  + \omegatil )x_{+} \wedge d\sigma^{+}
    +  x_{+ -} d\sigma^{-} \wedge d\sigma^{+}\,,   \\
     \xtil_{- +}d\sigma^{+} \wedge d\sigma^{-}
    &=& - (  -   \omega  + \omegatil )x_{-} \wedge d\sigma^{-}
    -  x_{- +} d\sigma^{+} \wedge d\sigma^{-}\,.   
\end{eqnarray*}
In principle we wish that the integrability conditions
$\xtil_{+-}=\xtil_{-+}$ are satisfied if the equations of motion
(\ref{eq:eom}) hold.  Subsequently we would like that this implies
equations of motion for $\xtil$.  We might as well substitute the
equations of motion for $x$ and $\xtil$ directly into the above and
find
\begin{eqnarray*}
    - \half \Htil_{kij} \xtil^{i}{}_{+} \xtil^{j}{}_{-} d\sigma^{-}
    \wedge d\sigma^{+} &=& + ( - \omega + \omegatil )_{ki}x^{i}{}_{+}
    \wedge d\sigma^{+} - \half H_{kij} x^{i}{}_{+} x^{j}{}_{-}
    d\sigma^{-} \wedge d\sigma^{+}\,, \\
     - \half \Htil_{kij} \xtil^{i}{}_{+} \xtil^{j}{}_{-} d\sigma^{+}
     \wedge d\sigma^{-} &=& - ( - \omega + \omegatil )_{kj}
     x^{j}{}_{-} \wedge d\sigma^{-} + \half H_{kij} x^{i}{}_{+}
     x^{j}{}_{-} d\sigma^{+} \wedge d\sigma^{-}\,.
\end{eqnarray*}
Next we selectively insert the pseudoduality equations (\ref{eq:special}) 
into the above
\begin{eqnarray}
    - \half \Htil_{kij} x^{i}{}_{+} \xtil^{j}{}_{-} d\sigma^{-}
    \wedge d\sigma^{+} &=& + ( - \omega + \omegatil )_{ki}x^{i}{}_{+}
    \wedge d\sigma^{+} 
    \nonumber \\
    & - & \half H_{kij} x^{i}{}_{+} x^{j}{}_{-}
    d\sigma^{-} \wedge d\sigma^{+}\,, 
    \label{eq:wedgeplus} \\
     + \half \Htil_{kij} \xtil^{i}{}_{+} x^{j}{}_{-} d\sigma^{+}
     \wedge d\sigma^{-} &=& - ( - \omega + \omegatil )_{kj}
     x^{j}{}_{-} \wedge d\sigma^{-} 
     \nonumber \\
     & + & \half H_{kij} x^{i}{}_{+}
     x^{j}{}_{-} d\sigma^{+} \wedge d\sigma^{-}\,.
     \label{eq:wedgeminus}
\end{eqnarray}
Let us first concentrate on the first equation above. We can choose 
$x^{i}{}_{+}$ to be arbitrary at any $\sigma$ so we conclude that
$$
    - \half \Htil_{kij}  \xtil^{j}{}_{-} d\sigma^{-}
    \wedge d\sigma^{+} = + (\omegatil  - \omega  )_{ki}
    \wedge d\sigma^{+} - \half H_{kij}  x^{j}{}_{-}
    d\sigma^{-} \wedge d\sigma^{+}\,.
$$
Next we substitute $(\omega^{j} - x^{j}{}_{+}d\sigma^{+})$ for
$x^{j}{}_{-} d\sigma^{-}$ and similarly for $\xtil^{j}{}_{-}
d\sigma^{-}$. This leads to
$$
    \left( \omegatil_{ki}-\omega_{ki} + \half \Htil_{kij}\omegatil^{j}
    - \half H_{kij}\omega^{j}\right) \wedge d\sigma^{+} =0\,.
$$
We see that there exists a tensor $U_{ki+}$, antisymmetric under $k 
\leftrightarrow i$, such that
\begin{equation}
    \left( \omegatil_{ki}-\omega_{ki} + \half \Htil_{kij}\omegatil^{j}
    - \half H_{kij}\omega^{j}\right) 
    = U_{ki+}d\sigma^{+}\,.
    \label{eq:defUplus}
\end{equation}
Next we concentrate on (\ref{eq:wedgeminus}) and observe that 
$x^{j}_{-}$ may be chosen arbitrarily at any $\sigma$. This leads to 
$$
     + \half \Htil_{kij} \xtil^{i}{}_{+} d\sigma^{+}
     \wedge d\sigma^{-} 
     = - (\omegatil  - \omega)_{kj} \wedge d\sigma^{-} 
     +  \half H_{kij} x^{i}{}_{+}
     d\sigma^{+} \wedge d\sigma^{-}\,.
$$
Next we substitute $(\omega^{i} - x^{i}{}_{-}d\sigma^{-})$ for 
$x^{i}{}_{+}d\sigma^{+}$ and similarly for 
$\xtil^{i}{}_{+}d\sigma^{+}$ with result
$$
    \left( \omegatil_{ki}-\omega_{ki} - \half \Htil_{kij}\omegatil^{j}
    + \half H_{kij}\omega^{j}\right) \wedge d\sigma^{-} =0\,.
$$
We see that there exists a tensor $U_{ki-}$, antisymmetric under $k 
\leftrightarrow i$, such that
\begin{equation}
    \left( \omegatil_{ki}-\omega_{ki} - \half \Htil_{kij}\omegatil^{j}
    + \half H_{kij}\omega^{j}\right) 
    = U_{ki-}d\sigma^{-}\,.
    \label{eq:defUminus}
\end{equation}
Adding and subtracting (\ref{eq:defUplus}) and (\ref{eq:defUminus}) we 
see that
\begin{eqnarray}
    \omegatil_{ki}-\omega_{ki} & = & 
    \half( U_{ki+}d\sigma^{+} + U_{ki-}d\sigma^{-})\,.
    \label{eq:sum}  \\
    \half \Htil_{kij}\omegatil^{j}
    - \half H_{kij}\omega^{j} & = & 
    \half( U_{ki+}d\sigma^{+} - U_{ki-}d\sigma^{-})\,.
    \label{eq:diff}
\end{eqnarray}
The latter equation above may be solved by substituting (\ref{eq:defvel}) 
and finding 
\begin{eqnarray}
    U_{ki+}d\sigma^{+} + U_{ki-}d\sigma^{-} & = & 
    \Htil_{kij}(\xtil^{j}{}_{+}d\sigma^{+} - \xtil^{j}{}_{-} d\sigma^{-})
    \nonumber  \\
     & - & H_{kij}(x^{j}{}_{+}d\sigma^{+} - x^{j}{}_{-}d\sigma^{-})\,,
    \nonumber  \\
     & = & \Htil_{kij}\omega^{j} - H_{kij}\omegatil^{j}\,.
    \label{eq:solU}
\end{eqnarray}
To obtain the bottom equation we used the duality relation (\ref{eq:special}) 
and also (\ref{eq:defvel}). We conclude that
\begin{equation}
    \omegatil_{ki} -\omega_{ki} = \half \Htil_{kij}\omega^{j} 
    - \half H_{kij}\omegatil^{j}\,.
    \label{eq:connection}
\end{equation}
These Pfaffian equations are the central result of this paper. They 
are the basic integrability condition for the pseudoduality 
equations \eqref{eq:special}. We will later discuss specific results that 
follow from  applying them to a variety of examples.
These Pfaffian equations along with \eqref{eq:derx} and the 
corresponding ``tilded'' equations should be viewed as defining a
distribution\footnote{Here we use distribution in the differential
geometric sense, see \cite{Warner}.  Said succinctly, a
$k$-dimensional distribution on a manifold $N$ is a rank $k$
sub-bundle of the tangent bundle $TN$, \emph{i.e.}, a $k$-plane field
on $N$.} in $\Sigma \times (\Orthframe(M)\times\Orthframe(\Mtil))$. 
The statement that the coordinates $(\sigma^{+},\sigma^{-})$ on
$\Sigma$ are the independent variables tells us that we should look
for integrable $2$-dimensional distributions that are solutions of the
above where $d\sigma^{-}\wedge d\sigma^{+}$ does not vanish when
restricted to this $2$-dimensional distribution.

\section{Integrability Conditions}
\label{sec:integrability}
Next we look for the conditions on the distribution defined by
\eqref{eq:connection} that allow for integrable $2$ dimensional
manifolds (worldsheets) when the equations of motion hold.  Taking the
exterior derivative of \eqref{eq:connection} and using the Cartan
structural equations leads to the following
\begin{equation}
    \nablatil_{k}\Htil_{lij} + \nablatil_{l}\Htil_{kij} =
    - \left( \nabla_{k} H_{lij} + \nabla_{l} H_{kij} \right)\,,
    \label{eq:HHtil}
\end{equation}
\begin{eqnarray}
     \Rtil_{ijkl} &-& \half \Htil_{ijm}\Htil_{mkl}
     -\frac{1}{4}\left( \Htil_{imk}\Htil_{mjl} - \Htil_{iml} 
     \Htil_{mjk}\right)
    \nonumber  \\
     & = & -\left[ R_{ijkl} - \half H_{ijm} H_{mkl}
     -\frac{1}{4}\left( H_{imk}H_{mjl} - H_{iml} H_{mjk}
     \right)\right]\,.
    \nonumber
\end{eqnarray}
In the above $\nabla$ and $\nablatil$ are respectively the covariant 
derivatives with respect to the riemannian connections $\omega_{ij}$ 
and $\omegatil_{ij}$. The reader is reminded that since $H$ and 
$\Htil$ are closed $3$-forms we have
\begin{eqnarray}
    \nabla_{i}H_{jkl} - \nabla_{j}H_{kli} + \nabla_{k}H_{lij} -
    \nabla_{l}H_{ijk} & = & 0\,,
    \label{eq:curlH}  \\
    \nablatil_{i}\Htil_{jkl} - \nablatil_{j}\Htil_{kli} +
    \nablatil_{k}\Htil_{lij} - \nablatil_{l}\Htil_{ijk} & = & 0\,.
    \label{eq:curlHtil}
\end{eqnarray}
Combining (\ref{eq:HHtil}) with the two equations above leads to the 
conclusion 
$$
    \nablatil_{i}\Htil_{jkl} = -\nabla_{i}H_{jkl} \,.
$$

Summarizing we see that the integrability equations for solving the
Pfaffian equations (\ref{eq:connection})
\begin{equation}
    \omegatil_{ij} -\half \Htil_{ijk}\omega^{k}
    = \omega_{ij} - \half H_{ijk} \omegatil^{k}\,.
    \label{eq:connection1}
\end{equation}
are
\begin{eqnarray}
     \Rtil_{ijkl} &-& \half \Htil_{ijm}\Htil_{mkl}
     -\frac{1}{4}\left( \Htil_{imk}\Htil_{mjl} - \Htil_{iml} 
     \Htil_{mjk}\right)
    \nonumber  \\
     & = & -\left[ R_{ijkl} - \half H_{ijm} H_{mkl}
     -\frac{1}{4}\left( H_{imk}H_{mjl} - H_{iml} H_{mjk}
     \right)\right]\,,
    \label{eq:RRtil} \\
    \nablatil_{i}\Htil_{jkl} & = & -\nabla_{i}H_{jkl} \,,
    \label{eq:HHtilsol}
\end{eqnarray}
and possible new integrability conditions found by taking the exterior
derivatives of the integrability equations above.  Notice that the 
right hand side of the above is not the curvature of a connection with 
torsion, see Appendix~\ref{sec:torsion}.

In the first paper \cite{Alvarez:2000pk} the condition that the spaces
be symmetric spaces arose from differentiating the above.  Here it is
convenient to define a tensor $S$ by
\begin{equation}
    S_{ijkl} = R_{ijkl} - \half H_{ijm} H_{mkl}
     -\frac{1}{4}\left( H_{imk}H_{mjl} - H_{iml} H_{mjk}\right)\,,
    \label{eq:defS}
\end{equation}
and similarly $\Stil$. The covariant differential of $S$ is given by
\begin{equation*}
    \nabla S = dS + \omega_{\bullet\bullet}\diamond S\,,
\end{equation*}
where $\omega_{\bullet\bullet}$ is an abstract notation for the 
$\mathfrak{s}\mathfrak{o}(n)$-valued connection $2$-form and $\diamond$ 
denotes the action of $\mathfrak{s}\mathfrak{o}(n)$ on $S$. A brief  
computation shows that
\begin{equation}
    \begin{split}
	\nabla_{k}S = \half \Htil_{\bullet\bullet k}\diamond \Stil\,,  \\
	\nablatil_{k}\Stil = \half H_{\bullet\bullet k}\diamond S\,.
    \end{split}
    \label{eq:derS}
\end{equation}

If $H=\Htil=0$ then $S=R$, $\Stil=\Rtil$ then we recover the symmetric
space conditions $\nabla R=0$ and $\nablatil \Rtil=0$, and the
opposite curvature conditions discussed in \cite{Alvarez:2000pk}. 
There we saw that dual symmetric spaces \cite{ONeill:SRG,Wolf:CC} gave
a class of manifolds with opposite curvature.  An interesting
mathematical question is suggested by our discussion.  Is there a
generalization of dual symmetric spaces that provides a framework for
the integrability conditions discussed in this Section?

\subsection{Relation to the Renormalization Group}
\label{sec:RG}

We make a brief remark about duality and the renormalization group. 
The first person to study this issue was Nappi \cite{Nappi:1980ig}
within the context of the Zakharov-Mikhailov model.  The connection
between off shell duality and the renormalization group was first
studied by Buscher \cite{Buscher:1988qj, Buscher:1988xu}.  Define a
tensor $S_{jk}$ by $S_{jl}=S_{ijil}$.  A brief computation shows that
\begin{equation*}
    S_{jl} = R_{jl} -\frac{1}{4}\, H_{imj}H_{iml}\,.
\end{equation*}
The $1$-loop renormalization group beta function \cite{Friedan:1980jf,
Curtright:1984dz} for the metric $g_{jl}$ is precisely $S_{jl}$. 
Similarly you notice that if in the other integrability condition
\eqref{eq:HHtilsol} you take a trace on $ik$ you get
$\nabla_{i}H_{jil}$ which is the $1$-loop beta function for the
$2$-form $B_{jl}$.  It was pointed out in \cite{Zachos:1994fa} that
the opposite signs in the beta functions found by Nappi
\cite{Nappi:1980ig} in the pseudodual models of Zakharov and Mikhailov
\cite{Zakharov:1978pp} are due to opposite signs of the generalized
curvatures \cite{Braaten:1985is}.  Here we have shown a more general
result.  Direct consequences of the integrability conditions for
pseudoduality \eqref{eq:RRtil} and \eqref{eq:HHtilsol} are that the
$1$-loop beta functions will have have opposite signs for any two
sigma models that are classically pseudodual.  Clearly there is some
interesting geometry in the space of field theories that is not yet
understood.

\section{Some Simple Examples}
\label{sec:examples}

We show that two well known dual models correspond to simple solutions
of (\ref{eq:connection}).  The best way to see this is to choose local
coordinates $(x,R_{V})$ and $(\xtil,\Rtil_{\Vtil})$ respectively on
$\Orthframe(M)$ and $\Orthframe(\Mtil)$.  We will look for solutions
that have $R_{V}=\Rtil_{\Vtil}=I$.  In this case equations
(\ref{eq:globalconn}) and (\ref{eq:conncoeff}) tell us that
\begin{eqnarray}
    \omega_{ij} & = & (\omega_{V})_{ij} = (\omega_{V})_{ijk} 
    \omega^{k}_{V}\,,
    \label{eq:omegatriv}  \\
    \omegatil_{ij} & = & (\omegatil_{\Vtil})_{ij} = 
    (\omegatil_{\Vtil})_{ijk} \omegatil^{k}_{\Vtil}\,.
    \label{eq:omegatiltriv}
\end{eqnarray}
In all of Section~\ref{sec:examples} we will work on the base
manifolds $M$ and $\Mtil$ and so we drop the $V$ and $\Vtil$
subscripts.  Inserting the above into (\ref{eq:connection}) leads to
\begin{equation}
    \omegatil_{ijk} \omegatil^{k} + \half H_{ijk}\omegatil^{k}
    = \omega_{ijk}\omega^{k} + \half \Htil_{ijk}\omega^{k}\,.
    \label{eq:conntriv}
\end{equation}
The hypotheses and the duality equations tell us that
$$
    \omega^{i} = x^{i}{}_{+}d\sigma^{+} + x^{i}{}_{-}d\sigma^{-}
    \quad\mbox{and}\quad
    \omegatil^{i} = x^{i}{}_{+}d\sigma^{+} - x^{i}{}_{-}d\sigma^{-}\,.
$$
Since $x^{i}{}_{+}$ and $x^{i}{}_{-}$ may be independently chosen at 
any point $\sigma$ we can \emph{de facto} treat $\omega^{i}$ and 
$\omegatil^{j}$ as being independent for our purposes. In this way we 
conclude that
\begin{eqnarray}
    \omegatil_{ijk} & = & - \half H_{ijk}\,,
    \label{eq:trivsol1}  \\
    \omega_{ijk} & = & -\half \Htil_{ijk}\,.
    \label{eq:trivsol2}
\end{eqnarray}

\subsection{Pseudochiral Model}
\label{sec:ZMmodel}

Here we discuss the pseudochiral model \cite{Zakharov:1978pp} of
Zakharov and Mikhailov.  Consider a sigma model with target space $M$ a
real connected compact Lie group $G$ with an $\Ad(G)$-invariant metric. 
The structure constants $f_{ijk}$ are skew symmetric, see
Appendix~\ref{sec:Liegroups}, and the coefficients of the riemannian
connection are given by $\omega_{ijk}= -\half f_{ijk}$.  This sigma
model also has $H_{ijk}=0$.  Applying this to (\ref{eq:trivsol1}) and
(\ref{eq:trivsol2}) we see that in the dual sigma model
$\omegatil_{ijk}=0$ and $\Htil_{ijk}=f_{ijk}$.  Since the connection
is trivial, the Cartan structural equation (\ref{eq:cartan1}) pulled
back to $\Mtil$ tells us that $d\omegatil^{i}=0$ and therefore we can
find coordinates so that $\omegatil^{i}= d\xtil^{i}$.  The trivialness
of the connection tells us that we can choose the manifold $\Mtil$ to
be euclidean space $\mathbb{R}^{n}$ that can be identified with 
the Lie algebra $\lieg$ of $G$.  Note that the $3$-form
$\Htil = \frac{1}{3!} f_{ijk} d\xtil^{i}\wedge d\xtil^{j} \wedge
d\xtil^{k}$ is closed as required.

\subsection{WZW Type Models}
\label{sec:WZWmodel}

In this case we take the sigma model $(M,g,B)$ to be a connected
compact real Lie group with an $\Ad(G)$-invariant metric.  The
Maurer-Cartan equations are \eqref{eq:MC}.  The $3$-form $H$ is taken
to be proportional to the structure constants $H_{ijk} = a f_{ijk}$
where $a \in \mathbb{R}$ is constant.  What is strictly called the WZW
model corresponds to $a=\pm 1$ with a specific normalization of the
action needed to make the path integral well defined.  Note that
worldsheet parity takes $a$ to $-a$ so we can restrict
ourselves to $a \ge 0$.  To work out the pseudodual sigma model we
insert the above into (\ref{eq:trivsol1}) and (\ref{eq:trivsol2})
where we find that $\omegatil_{ij} = - \half a f_{ijk}\omegatil^{k}$
and $\Htil_{ijk} = f_{ijk}$.  By using the first Cartan structural
equation we obtain the Maurer-Cartan equations
\begin{equation}
    d\omegatil^{i} = -\half a f_{ijk} \omegatil^{j} \wedge \omegatil^{k}\,.
    \label{eq:MC1}
\end{equation}
The dual manifold $\Mtil$ is the group $G$ because the Maurer-Cartan
equations above are just a rescaled version of \eqref{eq:MC}.  Note
that the metric on $(\Mtil,\gtil,\Btil)$ is $\gtil =
\omegatil^{i}\otimes \omegatil^{i}$ and the Maurer-Cartan equations
are \eqref{eq:MC1}.  The connection $\omegatil_{ij}$ must be the
riemannian connection for metric $\gtil$ so the metric $\gtil$ is a
rescaled version of the metric $g$ as we will see.  The $3$-form
$\Htil = \frac{1}{3!}f_{ijk} \omegatil^{i} \wedge\omegatil^{j}
\wedge\omegatil^{k}$ is closed as required\footnote{You can verify
that the geometric data satisfies the integrability conditions derived
in Section~\ref{sec:integrability} though it is not necessary to do so
in this case.}.  The model with $a=1$ is self pseudodual.  Also we
note that the $a\to 0$ limit of the dual model is the pseudochiral
model \cite{Ivanov:1987yv,Zachos:1994fa}.

There are a few observations worth making about the classical 
lagrangian. Classically, the overall normalization of the lagrangian 
is irrelevant. Schematically we can write the lagrangian for 
$(M,g,B)$ as 
\begin{equation*}
    \mathcal{L} = \omega^{i} \otimes \omega^{i} + a f_{ijk}\omega 
    ^{i}\wedge\omega^{j}\wedge \omega^{k}\,.
\end{equation*}
The lagrangian for $(\Mtil,\gtil,\Btil)$ is
\begin{equation*}
    \Tilde{\mathcal{L}} = \omegatil^{i} \otimes \omegatil^{i} + f_{ijk}\omegatil 
    ^{i}\wedge\omegatil^{j} \wedge\omegatil^{k}\,.
\end{equation*}
If we define $\omegahat^{i} = a \omegatil^{i}$ then $\omegahat$ 
satisfies the ``original'' Maurer-Cartan equations
\begin{equation}
    d\omegahat^{i} = -\half  f_{ijk} \omegahat^{j} \wedge \omegahat^{k}\,.
    \label{eq:MC2}
\end{equation}
and we can write the lagrangian as
\begin{equation*}
    \Tilde{\mathcal{L}} = \frac{1}{a^{2}} \left( \omegahat^{i} \otimes
    \omegahat^{i} + \frac{1}{a}\, f_{ijk}\omegahat ^{i}\wedge\omegahat^{j}
    \wedge\omegahat^{k}\right)\,.
\end{equation*}
At the level of equations of motion the pseudoduality transformation
takes the model with parameter $a$ to the one with parameter $1/a$. 
This result should be in the literature but I have not found an
explicit reference to it.  One final remark, it is a well known result
in differential geometry that rescaling the metric does not change the
connection $1$-form; you can verify that
$\omegatil_{ij}=\hat{\omega}_{ij}$.

\subsection{Explicit Computation in WZW Type Models}
\label{sec:WZWexplicit}

We can actually be very explicit and see how it all develops.  The
equations of motion for the WZW type model on $G$ with parameter $a$ can be
written as
\begin{equation}
    \partial_{-}\left(g^{-1}\partial_{+}g \right)
    + \partial_{+}\left(g^{-1}\partial_{-}g \right)
    = - a \left[g^{-1}\partial_{+}g \,, g^{-1}\partial_{-}g
    \right]_{G}\,.
    \label{eq:eomWZW}
\end{equation}
We put a subscript $G$ to identify the group associated with that 
Lie bracket. The equations of motion on $\Gtil$ with parameter 
$\tilde{a}$ are
\begin{equation}
    \partial_{-}\left(\gtil^{-1}\partial_{+}\gtil \right) +
    \partial_{+}\left(\gtil^{-1}\partial_{-}\gtil \right) = -
    \tilde{a} \left[\gtil^{-1}\partial_{+}\gtil \,,
    \gtil^{-1}\partial_{-}\gtil \right]_{\Gtil}\,.
    \label{eq:eomWZWtil}
\end{equation}
The general theory requires that we work with orthonormal frames. 
We choose an orthonormal basis $\{X_{i}\}$ for the Lie algebra of
$G$.  In this basis, the Lie brackets are given by $[X_{j},X_{k}]_{G}
= f^{i}{}_{jk}X_{i}$.  Similarly in $\Gtil$ we choose an orthonormal
basis $\{\Xtil_{i}\}$ with Lie brackets $[\Xtil_{j},\Xtil_{k}]_{\Gtil}
= \ftil^{i}{}_{jk}\Xtil_{i}$.  The duality equations are
\begin{align}
	\left(\gtil^{-1}\partial_{+}\gtil\right)^{i} &=
	+ \left( g^{-1}\partial_{+}g \right)^{i}\,, 
	\label{eq:WZWdual1} \\
	\left(\gtil^{-1}\partial_{-}\gtil\right)^{i}
	&= - \left(g^{-1}\partial_{-}g\right)^{i}\,.
    \label{eq:WZWdual2}
\end{align}
Subtract $\partial_{+}$ of \eqref{eq:WZWdual2} from $\partial_{-}$ 
of \eqref{eq:WZWdual1} to obtain
\begin{align*}
    \left[ \gtil^{-1}\partial_{+}\gtil\,, 
    \gtil^{-1}\partial_{-}\gtil \right]^{i}_{\Gtil} &=
    -a \left[ g^{-1}\partial_{+}g\,, g^{-1}\partial_{-}g 
    \right]^{i}_{G}\,,\\
\intertext{or}
    \ftil^{i}{}_{jk} \left(\gtil^{-1}\partial_{+}\gtil\right)^{j}
    \left(\gtil^{-1}\partial_{-}\gtil \right)^{k}
    &= -a f^{i}{}_{jk}\left(g^{-1}\partial_{+}g\right)^{j}
   \left(g^{-1}\partial_{-}g\right)^{k}\,.
\end{align*}
In deriving the above we only had to only use the equations of motion 
for $g$ not the equations of motion for $\gtil$. By using the duality 
relations we learn that
\begin{equation}
    \ftil_{ijk}= a f_{ijk}
    \label{eq:ftilaf}
\end{equation}
in agreement with \eqref{eq:MC1}. We can also consider the sum of
$\partial_{-}$ of \eqref{eq:WZWdual1} and $\partial_{+}$ of 
\eqref{eq:WZWdual1} to obtain
\begin{align}
        \partial_{-}\left( \gtil^{-1}\partial_{+}\gtil\right)^{i}
        +\partial_{+} \left( \gtil^{-1}\partial_{-}\gtil \right)^{i}
	&= \left[ g^{-1}\partial_{+}g\,, g^{-1}\partial_{-}g 
    \right]^{i}_{G}  \nonumber \\
        &= f^{i}{}_{jk} \left(g^{-1}\partial_{+}g\right)^{j}
   \left(g^{-1}\partial_{-}g\right)^{k} \,, \nonumber\\
   & = -f_{ijk} \left(\gtil^{-1}\partial_{+}\gtil\right)^{j}
    \left(\gtil^{-1}\partial_{-}\gtil \right)^{k}\,,
    \label{eq:WZWa1} \\
    & = - \frac{1}{a} \ftil_{ijk} \left(\gtil^{-1}\partial_{+}\gtil\right)^{j}
    \left(\gtil^{-1}\partial_{-}\gtil \right)^{k}\,,
    \nonumber \\
    & = - \frac{1}{a} \left[ 
    \gtil^{-1}\partial_{+}\gtil \,,
    \gtil^{-1}\partial_{-}\gtil \right]^{i}_{\Gtil}\,.
    \label{eq:WZWa2}
\end{align}
These are the equations of motion for the model on $\Gtil$.  We used
\eqref{eq:ftilaf} that depends only on the equations of motion of $g$,
and the duality relations \eqref{eq:WZWdual1}, \eqref{eq:WZWdual2}. 
Equations \eqref{eq:WZWa1} are the statement that $\Htil_{ijk} = f_{ijk}$.  We
showed that the equations of motion \eqref{eq:WZWa2} for $\gtil$ are
\eqref{eq:eomWZWtil} with $\tilde{a}=1/a$.

\section{Strict WZW Models}
\label{sec:strictWZW}

This example is generalizes the examples in
Section~\ref{sec:examples}.  There we solved \eqref{eq:conntriv} by
requiring $R_{V}=I$ and wrote an explicit solution on the base. 
Here we affirm that there are other other solutions when $R_{V}\neq
I$.  This is similar to the situation discussed in \cite[Section
2]{Alvarez:2000pk} where we saw that there were no pseudoduality
solutions if $T=I$ but there are solutions if we allowed $T$ to be an
orthogonal matrix. We find the very surprising result that any two strict 
WZW model on compact Lie groups of the same dimensionality are 
pseudodual.

Let $M=G$ be a compact connected Lie group of dimension $n$ with an
$\Ad(G)$-invariant metric.  Essentially what we want to do is choose
$H_{ijk}$ to be $a f_{ijk}$.  We have to be careful because $H$ is
defined on the orthonormal frame bundle of $G$ while the structure
constants are defined on $G$.  Since a Lie group is parallelizable we
choose a global orthonormal coframe $\omega_{V}^{i}$.  Note that the
open set $V$ is $G$.  The orthonormal frame bundle is trivial so
$\Orthframe(G) = G \times \SOrth(n)$.  At the point $(g,R_{V}) \in
\Orthframe(G)$ we define the functions $H_{ijk}$ by $H_{ijk} = a
(R_{V})_{il}(R_{V})_{jm}(R_{V})_{kn}f_{lmn}$ where $|a|=1$.  The
adjoint bundle\footnote{The adjoint bundle is the trivial bundle
$G\times \Ad(G)$ where $\Ad(G) \subset \SOrth(n)$ is the adjoint group
of $G$.} of $G$ is a sub-bundle of $\Orthframe(G)$ and the functions
$H_{ijk}$ restricted to the adjoint bundle are constant functions
given by $a f_{ijk}$.  Pulling back the right hand sides of
(\ref{eq:RRtil}) and (\ref{eq:HHtilsol}) to the base $G$ you see that
Appendix~\ref{sec:Liegroups} immediately tells you that they vanish. 
Choose $\Mtil=\Gtil$ to be any $n$-dimensional compact Lie group with
an $\Ad(\Gtil)$-invariant metric and with
$\Htil_{ijk}=\tilde{a}(\Rtil_{\Vtil})_{il}(\Rtil_{\Vtil})_{jm}(\Rtil_{\Vtil})_{kn}
\ftil_{lmn}$ where $|\tilde{a}|=1$.  The integrability conditions
(\ref{eq:RRtil}) and (\ref{eq:HHtilsol}) are trivially satisfied
(there are no further integrability conditions because $d0=0$) and
there are integrable $2$-dimensional distributions that solve
(\ref{eq:connection1}).  The WZW model on $G$ is pseudodual to the WZW
model on $\Gtil$ for any two compact $n$-dimensional Lie groups.  Note
that $\Gtil$ may be taken to be abelian\footnote{In the case of an
abelian group we do not have to worry about compactness since we are
just looking at local properties of the PDEs.}.  All our conclusions follow from
local statements about the PDEs and we have not discussed global
constraints on pseudoduality.  In this section we used the freedom of
varying the orthogonal matrices $R_{V}$ and $\Rtil_{\Vtil}$ from point
to point in the frame bundles.  This is something we could not do in
the construction of Section~\ref{sec:examples}.

\subsection{An Example}
\label{sec:example-abelian}

Here we work out explicitly the case of pseudoduality between a strict
WZW sigma model on a compact Lie group and the sigma model on an
abelian group of the same dimensionality.  We take $M$ to be an
abelian group such as $\bbR^{n}$ or $\mathbb{T}^{n}$.  The equations
of motion are $\partial^{2}_{+-}\phi^{i}=0$.  We take $\Mtil = \Gtil$
a compact Lie group with an $\Ad(G)$-invariant metric.  Let
$\{\Xtil_{i}\}$ be an orthonormal basis for the Lie algebra of $\Gtil$
with bracket relations $[\Xtil_{i},\Xtil_{j}]_{\Gtil} =
\ftil^{k}{}_{ij}\Xtil_{k}$.  The structure constants with lowered
indices $\ftil_{ijk}$ are totally antisymmetric in $ijk$, see
Appendix~\ref{sec:Liegroups}.  In a strict WZW model the equations of
motion may be written as $\partial_{-}(\gtil^{-1}\partial_{+}\gtil)=0$
where $\gtil:\Sigma \to \Mtil$.  The pseudoduality equations are
\begin{equation}
    \begin{split}
        (\gtil^{-1}\partial_{+}\gtil)^{i} &= 
        +T^{i}{}_{j}\partial_{+}\phi^{j}\,,  \\
        (\gtil^{-1}\partial_{-}\gtil)^{i} &= 
        -T^{i}{}_{j}\partial_{-}\phi^{j}\,,      
    \end{split}
    \label{eq:psdab}
\end{equation}
where $T$ is an orthogonal matrix and $\gtil^{-1}d\gtil =
(\gtil^{-1}d\gtil)^{i} \Xtil_{i}$.  Taking $\partial_{-}$ of the first
equation above we learn that $(\partial_{-}T)(\partial_{+}\phi)=0$. 
Since we can choose $\partial_{+}\phi$ to have an arbitrary value at
any $\sigma$ we have that $\partial_{-}T=0$.  Thus we learn that $T$
is a function of only $\sigma^{+}$.  Next we take $\partial_{+}$ of
the second equation above, use the equations of motion and conclude
that
\begin{equation}
    \left[(\partial_{+}T)T^{-1}\right]^{i}{}_{j}
    = -\ftil^{i}{}_{kj}T^{k}{}_{l}\partial_{+}\phi^{l}\,.
    \label{eq:psdab1}
\end{equation}
We note that the right hand side is skew under $i\leftrightarrow j$
and is only a function of only $\sigma^{+}$.  Therefore, we have an
ordinary differential equation \eqref{eq:psdab1} that will produce an
orthogonal matrix $T(\sigma^{+})$ that depends on
$\partial_{+}\phi(\sigma^{+})$.

Summarizing we have seen that for any solution $\phi^{i}$ of the wave 
equation we can construct an orthogonal matrix $T$ and subsequently 
use \eqref{eq:psdab} to construct solutions to the strict WZW model 
on a compact simple Lie group.

The reader could ask whether we worked too hard in this section.  The
equations of motion tell us that $\gtil^{-1}\dplus \gtil$ and
$\dplus\phi$ and only functions of $\sigma^{+}$.  Does it not suffice
to use only the first of \eqref{eq:psdab} and any arbitrary
$T(\sigma^{+})$, not necessarily orthogonal, and in that way map a
solution of the free equation into a solution of the strict WZW model? 
There is a reason for invoking the second equation in
\eqref{eq:psdab}.  It is desirable to preserve the stress energy 
tensor. The construction described in this paragraph will preserve 
$\Theta_{++}$ if $T$ is orthogonal. Anything can happen to 
$\Theta_{--}$. By requiring both equations in \eqref{eq:psdab} we are 
guaranteeing that the stress-energy tensor is preserved. An analogous 
remark can be made in Section~\ref{sec:example-nonabelian}.

\subsection{A More Complicated Example}
\label{sec:example-nonabelian}

Here we consider the case where we consider pseudoduality between
strict WZW models where $M$ and $\Mtil$ are compact Lie groups of
dimension $n$ with $\Ad$-invariant metrics.  
Let $\{X_{i}\}$ be an orthonormal basis for the Lie algebra of $G$
with bracket relations $[X_{i},X_{j}]_{G} = f^{k}{}_{ij}X_{k}$.  The
structure constants $f_{ijk}$ are totally antisymmetric in $ijk$. 
Likewise we make analogous definitions for the Lie group $\Gtil$.  The
equations of motion are $\partial_{-}(g^{-1}\partial_{+}g) = 0$ and
$\partial_{-}(\gtil^{-1}\partial_{+}\gtil)=0$.  The pseudoduality
equations are
\begin{equation}
    \begin{split}
	(\gtil^{-1}\partial_{+}\gtil)^{i} &= 
	+T^{i}{}_{j}(g^{-1}\partial_{+}g)^{j}\,,  \\
	(\gtil^{-1}\partial_{-}\gtil)^{i} &= 
	-T^{i}{}_{j}(g^{-1}\partial_{-}g)^{j}\,,      
    \end{split}
    \label{eq:psdabx}
\end{equation}
where $T$ is an orthogonal matrix.  Taking $\partial_{-}$ of the first
equation above we learn that $(\partial_{-}T)=0$ and therefore $T$ is
a function of $\sigma^{+}$ only.  Taking $\partial_{+}$ of the second
equation above we learn that
\begin{equation}
    \begin{split}
    \left[(\partial_{+}T)T^{-1}\right]^{i}{}_{j}
    &= -\ftil^{i}{}_{kj}T^{k}{}_{l}(g^{-1}\partial_{+}g)^{l}
    + T^{i}{}_{k}T^{j}{}_{l}f_{kml}(g^{-1}\partial_{+}g)^{m}\,,\\
    &= \left( -\ftil_{imj}
    + T^{i}{}_{k} T^{m}{}_{p} T^{j}{}_{l} f_{kpl} \right)
    T^{m}{}_{n}(g^{-1}\partial_{+}g)^{n}\,.
    \end{split}
    \label{eq:psdabx1}
\end{equation}
In deriving the above we used $T^{-1}=T^{t}$. Note that the right 
hand side is skew under $i \leftrightarrow j$ and that everything on 
the right hand side is a function of $\sigma^{+}$ only. Thus the above 
is an ordinary differential equation with solution an orthogonal 
matrix $T(\sigma^{+})$.

Summarizing we have seen that for any solution $g$ of the equations of
motion for the strict WZW model on $G$ we can construct an orthogonal
matrix $T$ and subsequently use \eqref{eq:psdabx} to construct
a solution $\gtil$ to the strict WZW model on $\Gtil$. For example you 
could take the group $G$ to be $\SOrth(2l+1)$ associated with the Lie 
algebra $B_{l}$ and $\Gtil$ to be the compact symplectic group 
$\U_{\mathbb{H}}(l)$ associated with the Lie algebra $C_{l}$. 
Note that $\dim G = \dim \Gtil = l(2l+1)$.

We can make contact with the discussion in
Section~\ref{sec:WZWexplicit} with $a=1$ by noting that if $G=\Gtil$
then $T=I$ is a solution to \eqref{eq:psdabx1}.

\subsection{Some Geometry}
\label{sec:WZWgeometry}

We have compact Lie groups $G$ and $\Gtil$ of dimension $n$ with 
$\Ad$-invariant inner products on each. The adjoint action of the 
groups acts via isometries on the Lie algebras and therefore we can 
think of the respective adjoint groups $\Ad G$ and $\Ad \Gtil$ as 
subgroups of $\SOrth(n)$. We note that if we pick an orthonormal basis 
for the Lie algebra $\lieg$ then the structure constants are invariant 
under the adjoint action of $G$ and likewise for $\Gtil$ and 
$\liegtil$. There is a natural action of $\Ad G \times \Ad\Gtil$ on 
$T$ given by $(R,\Rtil) \in \Ad G \times \Ad\Gtil$ that takes $T$ into 
$\Rtil T R^{-1}$, see \eqref{eq:psdabx}. Since $f_{ijk}$ and 
$\ftil_{ijk}$ are respectively $\Ad G$ and $\Ad\Gtil$ invariant we 
have that differential equation \eqref{eq:psdabx1} is $\Ad G \times 
\Ad\Gtil$ invariant. When we parametrize our solutions as 
$\gtil(\sigma;g,T_{0})$ we see that we should really think of the 
solution as being parametrized by the equivalence class $[T_{0}] \in 
\Ad\Gtil\backslash \SOrth(n)/\Ad G$.

Finally since this section is supposed to be self-contained, I should
explain how to make sense of \eqref{eq:psdabx}.  After all, the right
hand side involves $\lieg$, the Lie algebra of $G$, while the left
hand side involves $\liegtil$, the Lie algebra of $\Gtil$.  Let
$\Isom(\lieg,\liegtil)$ be the vector space isometries from $\lieg$ to
$\liegtil$.  All we are saying is that we need a map $T:\Sigma \to
\Isom(\lieg,\liegtil)$ such that $\hodge (\gtil^{-1}d\gtil)(\sigma) =
T(g^{-1}dg)(\sigma)$, where $\hodge$ is the Hodge duality operator on
$\Sigma$.

We can even expand more on the above by rewriting \eqref{eq:psdabx1} 
is a different way
\begin{equation}
    \dplus T_{ij} = - [\ftil_{ilk}(\gtil^{-1}\dplus\gtil)^{l}]T_{kj}
    + T_{ik}[f_{klj}(g^{-1}\dplus g)^{l}] \,.
    \label{eq:psdabx2}
\end{equation}
The right hand side of this equation is Lie algebra version of the
$\Ad G \times \Ad \Gtil$ action on $T$.  It is straightforward to
solve this is equation but let us be a bit more abstract so that we
can state the solution in a coordinate independent fashion.  On
$\lieg$ define the adjoint action by $\ad_{\lieg}(X)Y = [X,Y]$ for
$X,Y\in\lieg$.  The vector space $\lieg$ has an inner product so we
can define $\ad_{\lieg}^{\dagger}: \lieg \to \lieg$ as the adjoint of
the transformation $\ad_{\lieg}$.  Since the metric on $\lieg$ is $\Ad
G$ invariant we have that $\ad_{\lieg}$ is a skew adjoint
transformation $\ad_{\lieg}^{\dagger} = - \ad_{\lieg}$.  The tangent
bundle of $G$ is trivial $TG = G \times \lieg$.  We have a map
$g:\Sigma \to G$ that can be used to pullback the tangent bundle to
$\Sigma$.  On this pullback bundle $g^{*}(TG)$ we define a flat
orthogonal connection by $\ad_{\lieg}(J^{(R)})$ where $J^{(R)} =
(g^{-1}\dplus g)d\sigma^{+}$.  This connection is flat by the
equations of motion and it is an orthogonal connection because
$\ad_{\lieg}$ is skew adjoint.  Let $P(\sigma)$ be parallel transport
from $(0,0)$ to $\sigma=(\sigma^{+},\sigma^{-})$.  Notice that since
$J^{(R)}$ is flat and it does not have a $d\sigma^{-}$ component we
have that $P(\sigma)$ is independent of $\sigma^{-}$.  We can define
similar structures on $\Gtil$ and $\gtil$.  From experience we know
that the integration of \eqref{eq:psdabx2} is given by parallel
transport.  Since one index of $T$ lives in $\lieg$ and the other in
$\liegtil$ we have that the solution of the equation above may be
written as
\begin{equation}
    T(\sigma) = \Ptil(\sigma)T_{0}P(\sigma)^{-1}\,,
    \label{eq:parallel}
\end{equation}
where $T(0)=T_{0}$.

This leads to a beautiful geometrical way to think about the
pseudoduality equations \eqref{eq:psdabx1}.  The equations of motion
tell us that there are natural flat connections $\ad_{\lieg}(J^{(R)})$
and $\ad_{\liegtil}(\Jtil^{(R)})$ respectively on the pullback bundles
$g^{*}(TG)$ and $\gtil^{*}(T\Gtil)$.  The solution of the ODE for T
tells us that the geometric content of pseudoduality is the following. 
Begin with $(g^{-1}dg)(\sigma)$ and parallel transport it to the
origin $P(\sigma)^{-1}(g^{-1}dg)(\sigma)$.  Do the same thing on the
dual model.  The fibers over the origin of the aforementioned bundles
are $\lieg$ and $\liegtil$.  Use a fixed isometry $T_{0}\in
\Isom(\lieg,\liegtil)$ and Hodge duality to equate these two
quantities\footnote{This resembles a result of \cite{Braaten:1985is}
for $\Gtil$ abelian but it is not.  The results here are duality based
and motivated while the observation of Braaten, Curtright and Zachos
is closely related to the discussion of Section~\ref{sec:detour} and
not directly related to pseudoduality.}:
\begin{equation}
    \hodge\left(\Ptil(\sigma)^{-1}(\gtil^{-1}d\gtil)\right) = 
    T_{0}\left(P(\sigma)^{-1}(g^{-1}dg)\right)\,.
    \label{eq:dualWZW}
\end{equation}
This equation is totally intrinsic without reference to bases,
\emph{etc.}, and encapsulates how pseudoduality transformation
operates on strict WZW models.  Note that in general you cannot use
the action of $\Ad G\times \Ad\Gtil$ to set $T_{0}=I$.

We would like to point out that the above is not the most convenient
approach from a computational viewpoint if you are looking for
pseudodual solutions.  To do this you begin with a $g(\sigma)$ and you
use \eqref{eq:psdabx1} to solve for $T$ and then integrate
\eqref{eq:psdabx} to find $\gtil(\sigma)$.

\subsection{Infinite Number of Conservation Laws}
\label{sec:conservation}

The discussions of Sections \ref{sec:example-abelian} and
\ref{sec:example-nonabelian} lead to a new method that can be used to
find an infinite number of non-local conserved currents by a variant
of a technique discussed in \cite{Pohlmeyer:1975nb}.  For a recent
discussion and references to the older literature on local and
non-local conservation laws for sigma models based on groups see
\cite{Evans:1999mj,Evans:2000hx}.  The connection between the method
described here and other methods is not clear.

We begin with the strict WZW model based on Lie group $G$ where the
basic local conserved currents are $J^{(R)} = (g^{-1}\dplus
g)d\sigma^{+}$ and $J^{(L)} = (\dminus g)g^{-1}d\sigma^{-}$.  It is
well known that powers of $J^{(R)}$ and $J^{(L)}$ give higher rank
local conservation laws.  What we would like to do is construct an
infinite number of non-local conservation laws on $G$ by using the
pseudodual model on a compact Lie group $\Gtil$ in an auxiliary
fashion.  We solve \eqref{eq:psdabx1} for $T$.  The ordinary
differential equation needs an initial condition $T(\sigma^{+}=0) =
T_{0} \in \SOrth(n)$.  To be more precise we note that $T =
T(\sigma^{+}; g,T_{0})$.  We use \eqref{eq:psdabx} to solve for $\gtil
= \gtil(\sigma; g,T_{0})$.  Next we construct the basic conserved
currents $\Jtil^{(R)}$ and $\Jtil^{(L)}$ on $\Gtil$.  We think of
$\Jtil^{(R)}$ and $\Jtil^{(L)}$ as functions of $g$ and $T_{0}$.  The
current $\Jtil^{(R)}$ is a non-local function of $g^{-1}\dplus g$
since \eqref{eq:psdabx1} and the first of \eqref{eq:psdabx} are
functions of $g^{-1}\dplus g$ only.  You get a family of non-local
conserved currents on the WZW model on $G$ parametrized by the initial
condition $T_{0}$.  If you write $T_{0} = e^{\alpha}$ where $\alpha$
is an antisymmetric matrix and you power series expand about
$\alpha=0$ then you will get an infinite number of non-local conserved
currents on the WZW model on $G$ (not $\Gtil$) that we can
schematically organize as $\Jtil^{(R)}(g,T_{0}) =
\sum_{n=0}^{\infty}\alpha^{n}\Jtil^{(R)}_{[n]}(g)$.  The first one
$\Jtil^{(R)}_{[0]}$ will be $J^{(R)}$ if $G=\Gtil$.  If $G \neq \Gtil$
then $T$ is nontrivial even for initial condition $T(0)=I$ and we 
cannot write down $\Jtil^{(R)}_{[0]}$ explicitly.  The other
current $\Jtil^{(L)}$ is more ``interesting'' because you need both
the equations in \eqref{eq:psdabx} to work out what it is.  You can do
a similar power series $\Jtil^{(L)}(g,T_{0}) =
\sum_{n=0}^{\infty}\alpha^{n}\Jtil^{(L)}_{[n]}(g)$ to get an infinite
number of non-local conserved currents.  Note that by choosing a
different group $\Gtil$ we get a different set of conservation laws
since $T$ depends on the choice of groups, see \eqref{eq:psdabx1}.

\section{Some Geometry of the Connections}
\label{sec:geometry}

In this section we study some of the geometry of the connections that
arises due to the Pfaffian equation \eqref{eq:connection1}.  In this
pursuit we run into an interesting fork in the road.  Motivated by
duality we obtain some results that are really properties of  sigma
models and do not have anything to do with duality.

In order to be very clear about what it happening it is convenient to
explicitly worry about pullbacks of differential forms.  On the bundle
of orthonormal frames $\Orthframe(M,g)$ defined by metric $g$ on $M$
we have a global coframing $(\omega^{i},\omega_{jk})$.  We also have a
map $X:\Sigma \to \Orthframe(M,g)$.  We can use $X$ to pull all
structures back to $\Sigma$ and so we get ``vector-valued $1$-forms''
$(\xi^{i},\xi_{jk})$ on $\Sigma$ defined by
\begin{eqnarray}
    X^{*}\omega^{i} & = & \xi^{i}\,,
    \label{eq:eta}  \\
    X^{*}\omega_{jk} & = & \xi_{jk}\,.
    \label{eq:etaconn}
\end{eqnarray}
By taking the exterior derivative and using (\ref{eq:cartan1}) and 
(\ref{eq:cartan2}) you find that the $\xi$ satisfy
\begin{eqnarray}
    d\xi^{i} & = & -\xi_{ij}\wedge\xi^{j}\,,
    \label{eq:etacartan1}  \\
    d\xi_{ij} & = & -\xi_{ik}\wedge\xi_{kj} +  \half
    r_{ijkl} \xi^{k}\wedge\xi^{l}\,,
    \label{eq:etacartan2}
\end{eqnarray}
where $r_{ijkl} = X^{*}R_{ijkl} = R_{ijkl}\circ X$ denotes the pullback
to $\Sigma$ of the functions $R_{ijkl}$ on $\Orthframe(M,g)$.

To write down the equations of motion we need the Hodge duality
operator $\hodge$ on $\Sigma$.  On $1$-forms it is given by
$\hodge(d\sigma^{\pm})= \pm d\sigma^{\pm}$.  For future reference we
note that if $\alpha$, $\beta$ are $1$-forms on $\Sigma$ then
\begin{equation}
    (\hodge \alpha)\wedge(\hodge\beta) = - 
    \alpha\wedge\beta\,,
    \label{eq:hodge}
\end{equation}
and that $(\hodge)^{2}\alpha=\alpha$.  Also note that $\alpha \wedge
(\hodge\beta) = \beta \wedge (\hodge\alpha)$.  In particular, you have 
that $\xi^{j}\wedge(\hodge\xi^{k})$ is symmetric under $j 
\leftrightarrow k$.
The sigma model is
specified by a map $X:\Sigma\to\Orthframe(M,g)$ that satisfies
\begin{eqnarray}
    d\xi_{ij}  +\xi_{ik}\wedge\xi_{kj} & = &  \half
    r_{ijkl} \xi^{k}\wedge\xi^{l}\,,
    \label{eq:eta2} \\
    d\xi^{i}  +\xi_{ij}\wedge\xi^{j}& = & 0\,,
    \label{eq:eta1}  \\
    d(\hodge \xi^{i}) + \xi_{ij}\wedge (\hodge \xi^{j}) &=&
    \half h_{ijk} \xi^{j}\wedge\xi^{k}\,,
    \label{eq:eta3}
\end{eqnarray}
where $h_{ijk} = X^{*}H_{ijk} = H_{ijk}\circ X$.  The last equation
above is the non-linear wave equation for the sigma model.

In an obvious notation, the pseudoduality equations are
\begin{equation}
    \xitil^{i} = \hodge \xi^{i}\,.
    \label{eq:psd3}
\end{equation}
Equation (\ref{eq:connection1}) may be written as
\begin{equation}
    \xitil_{ij} -\half \htil_{ijk}(\hodge\xitil^{k})
    = \xi_{ij} - \half h_{ijk} (\hodge\xi^{k})\,.
    \label{eq:connection2}
\end{equation}
Notice that everything on the left hand side refers to $\Mtil$ and
everything on the right hand side refers to $M$.  Motivated by the
equations of motion \eqref{eq:eom} and not pseudoduality, earlier
authors, see \emph{e.g.}, \cite{Braaten:1985is, Ivanov:1987yv},
suggested defining a ``connection'' by
\begin{equation}
    \xi'_{ij} = \xi_{ij} - \half h_{ijk} (\hodge\xi^{k})\,.
    \label{eq:psiconn}
\end{equation}
You have to be careful here for in general $\xi'_{ij}$ is not the
pullback of a connection on $\Orthframe(M,g)$ as we will see in the
next subsection; though $\xi'_{ij}$ is a connection on the pullback
bundle $X^{*}\Orthframe(M,g)$.

\subsection{Detour}
\label{sec:detour}

We now take a fork in the road and for the moment we forget about
$\Mtil$ and duality.  We try to rewrite the equations of motion for
the sigma model on $M$ in terms of $\xi'_{jk}$.  You find
\begin{eqnarray}
    d\xi'_{ij}  & + &\xi'_{ik}\wedge\xi'_{kj} = 
    -\half \left(\nabla^{\xi}_{k}h_{lij} + 
    \nabla^{\xi}_{l}h_{kij}\right) \xi^{k}\wedge (\hodge\xi^{l})
    \nonumber \\
     &+&  \half \left[ 
    r_{ijkl}  -  \half h_{ijm}h_{mkl}
    - \frac{1}{4}(h_{imk}h_{mjl} - h_{iml}h_{mjk}) \right]
    \xi^{k}\wedge\xi^{l}
    \,,
    \label{eq:eta2p} \\
    d\xi^{i}  & + &\xi'_{ij}\wedge\xi^{j} = 0\,,
    \label{eq:eta1p}  \\
    d(\hodge \xi^{i}) & + & \xi'_{ij}\wedge (\hodge \xi^{j}) =
    0 \,.
    \label{eq:eta3p}
\end{eqnarray}
The covariant derivative of $h_{ijk}$ is defined by
\begin{equation}
    \nabla^{\xi}h_{ijk} = dh_{ijk} + \xi_{il}h_{ljk}
    + \xi_{jl}h_{ilk} + \xi_{kl}h_{ijl} 
    = X^{*}(\nabla^{\omega}H_{ijk})\,.
    \label{eq:covh}
\end{equation}
Equations (\ref{eq:eta1p}) and (\ref{eq:eta3p}) look like the
equations (\ref{eq:eta1}) and (\ref{eq:eta3}) for a sigma model with
vanishing $3$-form.  Is there a lagrangian that gives these equations
of motion?  The affirmative answer requires that $\xi'_{ij}$ is the
pullback of a connection and that (\ref{eq:eta2p}) is of form
(\ref{eq:eta2}).  Let us be more precise.  Can we find a metric $g'$
on a new manifold $M'$ such that $\xi'_{ij}$ may be interpreted as the
pullback\footnote{If $M$ and $M'$ are spin manifolds then the pullback
bundle to $\Sigma$ of the respective spin frame bundles will be
trivial bundles and therefore isomorphic.  If $\Sigma$ has nontrivial
topology then you have to be careful if the manifolds are not spin
because $\pi_{1}(\SOrth(n)) = \mathbb{Z}_{2}$ for $n >2$.} to $\Sigma$
of a connection on $\Orthframe(M',g')$?  This bundle has a global
coframing $\{\theta^{i},\theta_{jk}\}$ that satisfies the Cartan
structural equations
\begin{eqnarray}
    d\theta^{i} & = & -\theta_{ij}\wedge \theta^{j} +
    \half T_{ijk} \theta^{j}\wedge\theta^{k}\,,
    \label{eq:tor1}  \\
    d\theta_{ij} & = & -\theta_{ik}\wedge\theta_{kj}
    + \half K_{ijkl}\theta^{k}\wedge\theta^{l},
    \label{eq:tor2}
\end{eqnarray}
where $T_{ijk}$ is the torsion of the connection $\theta_{ij}$ and
$K_{ijkl}$ is the curvature of the connection $\theta_{ij}$. The new sigma 
model is defined by a map $Y:\Sigma \to \Orthframe(M',g')$ satisfying
\begin{eqnarray*}
    Y^{*}\theta^{i} & = & \xi^{i}\,,  \\
    Y^{*}\theta_{ij} & = & \xi'_{ij}\,.
\end{eqnarray*}
Taking the exterior derivative of the first equation above and using 
(\ref{eq:eta1p}) leads to the conclusion that $T_{ijk}=0$. We have 
learned that the connection on $\Orthframe(M',g')$ is the unique 
torsion free riemannian connection associated to the metric $g'$. 
Taking the exterior derivative of the second equation above tells us 
that
\begin{eqnarray*}
        \half (Y^{*} K_{ijkl})\xi^{k}\wedge\xi^{l} &= &
    -\half \left(\nabla^{\xi}_{k}h_{lij} + 
    \nabla^{\xi}_{l}h_{kij}\right) \xi^{k}\wedge (\hodge\xi^{l})
    \\
     &+&  \half \left[ 
    r_{ijkl}  -  \half h_{ijm}h_{mkl}
    - \frac{1}{4}(h_{imk}h_{mjl} - h_{iml}h_{mjk}) \right]
    \xi^{k}\wedge\xi^{l}
    \,.
\end{eqnarray*}
Comparing both sides we learn that $\nabla^{\omega}_{k}H_{lij} +
\nabla^{\omega}_{l}H_{kij} =0$.  Combining this with $dH=0$, see
(\ref{eq:curlH}), tells us that $\nabla^{\omega}_{l}H_{ijk}=0$. Thus 
the full content of the equation above is
\begin{eqnarray}
    \nabla^{\omega}_{l}H_{kij}& = & 0 \,,
    \label{eq:fake1}  \\
    K_{ijkl} & = & R_{ijkl}
      -  \half H_{ijm}H_{mkl}
    - \frac{1}{4}(H_{imk}H_{mjl} - H_{iml}H_{mjk})\,.
    \label{eq:fake2}
\end{eqnarray}
The above should be viewed as equations on $\Orthframe(M,g)\times
\Orthframe(M',g')$.  Additional integrability conditions following
from taking derivatives of the above also have to be satisfied. 
Notice that the right hand side of \eqref{eq:fake2} is precisely the
tensor $S_{ijkl}$, see \eqref{eq:defS}.  Some authors have tried to
rewrite sigma model equations in terms of the orthogonal connection
with torsion $\omega_{ij}- \half H_{ijk}\omega^{k}$ on
$\Orthframe(M,g)$ but we are not big advocates of this because it does
not appear naturally in the geometrical framework, see \emph{e.g.},
the discussion above or \eqref{eq:connection}.  We feel that the
important relevant geometrical object is the pullback connection
\eqref{eq:psiconn} on the bundle $x^{*}\Orthframe(M,g)$.

The conclusion here is that the sigma model specified by geometric
data $(M,g,B)$ is equivalent to the sigma model $(M',g',B')$ with
$H'=dB'=0$ if the integrability equations above are satisfied. 
Equivalence is in the sense that there is a mapping that takes
solutions of sigma model $(M,g,B)$ into solutions of the other sigma
model $(M',g',B')$ and vice versa.  Notice that this part of the
discussion follows only from trying to identify \eqref{eq:psiconn} as
the pullback of a connection.  It is independent of the duality
motivation that lead to it. We are really discussing properties of 
sigma models and their equations of motion.

\subsection{Earlier Observations}
\label{sec:ivanov}

Braaten, Curtright and Zachos \cite{Braaten:1985is}
observed\footnote{These authors had a more restrictive Jacobi identity
condition on $H_{ijk}$, but not necessary, that was motivated by the
model they were studying.} that if the right hand sides of
\eqref{eq:fake1} and \eqref{eq:fake2} vanish then the manifold $\Mtil$
is $\bbR^{n}$ or $\mathbb{T}^{n}$.  They used the flatness of the
$\xi'_{ij}$ connection to solve the equations of motion in terms of a
free field and the parallel transport operator.  Our way of seeing
this is to observe that the connection $\theta_{ij}$ on $\Mtil$ is a
flat torsion free metric connection.

Ivanov \cite{Ivanov:1987yv} observed that if $M=G$ is a compact
semi-simple Lie group then the equations above have a solution. 
Choose an orthonormal global framing for $G$ and pull everything 
\eqref{eq:fake1} and \eqref{eq:fake2} back to $G$.
Assume that in metric $g$, the structure coefficients for the Lie group
in this orthonormal frame are given by $f_{ijk}$, see
Appendix~\ref{sec:Liegroups}.  Assume $H_{ijk} = b f_{ijk}$ where $b$
is a constant.  Then \eqref{eq:fake1} is automatically satisfied
because of \eqref{eq:covf}.  A brief computation shows that
\eqref{eq:fake2} is given by
\begin{equation}
    K_{ijkl} = \frac{1}{4}\,(1-b^{2}) f_{ijm}f_{mkl}\,.
    \label{eq:KLie}
\end{equation}
If we take a new metric $g'$ on $M'=G$ to be $g'=g/(1-b^{2})$, for $\lvert
b\rvert <1$, and $\theta_{ij}$ to be the torsion free riemannian
connection with respect to $g'$ then we are done.  This shows that
every solution to the the equations of motion for the generalized WZW
model on $G$ defined by metric $g$ and $H_{ijk}=b f_{ijk}$ may be
identified with a solution to the nonlinear sigma model on $G$ with
metric $g'=g/(1-b^{2})$ and $H'_{ijk}=0$.  We can now apply the
special case of pseudoduality discussed in
\cite{Alvarez:2000pk,Ivanov:1987yv}.  We know that the model on the
Lie group\footnote{The Lie group $G$ is viewed as the
symmetric space $G\times G/G$.} $G$ with $H'=0$ is pseudodual to a
model on the negative curvature symmetric space $\Mtil = G^{\mathbb{C}}/G$. 
Here $G^{\mathbb{C}}$ is the complexification of $G$.  So we see that
in the sense described above the generalized WZW model on $G$ with
$\lvert b\rvert <1$ is pseudodual to the model on $\Mtil =
G^{\mathbb{C}}/G$ with $\Htil_{ijk}=0$.

If $b=1$ in \eqref{eq:KLie} then $K_{ijkl}=0$ and we can take
$M'=\bbR^{n}$ or $M'=\mathbb{T}^{n}$ as noticed in
\cite{Braaten:1985is, Ivanov:1987yv}.

\subsection{Back to Pseudoduality}
\label{sec:back}

We briefly return to pseudoduality and make a few comments. We can 
mimic what was done on Section~\ref{sec:detour} with 
\eqref{eq:connection2} without introducing $\xi'_{ij}$ or 
$\tilde{\xi}'_{ij}$. We think of the left hand side and the right 
hand side of \eqref{eq:connection2} respectively as pullbacks of 
connections from the appropriate bundles. We will find that the 
compatibility conditions are precisely \eqref{eq:RRtil} and 
\eqref{eq:HHtilsol}. This method is mathematically equivalent to that 
used earlier in the article.

\section*{Acknowledgments}

I would like to thank T.~Curtright, L.~Ferreira, L.~Mezincescu,
R.~Nepomechie, J.~S\'{a}nchez Guill\'{e}n, I.M.~Singer, P.~Windey and
C.~Zachos for discussions.  This work was supported in part by
National Science Foundation grants PHY--9870101 and PHY--0098088.

\appendix

\section{$G$-structures}
\label{sec:g-structures}

$G$-structures arise as the imposition of a natural geometric structure 
on the tangent bundle of a manifold. The basic bundle in our discussion 
will be the {\em coframe\/} bundle $\coframe(M)$ of the manifold $M$. A 
point in $\coframe(M)$ consists of a point $x$ on the base manifold $M$ and a 
basis for the cotangent bundle at $x$. A coframe is a local section of 
this bundle. 
If $V\subset M$ is a neighborhood then we denote a coframe by
\[
\omega_V = \begin{pmatrix} \omega^1_V \\
    \omega^2_V \\ \vdots \\ \omega^n_V \\
    \end{pmatrix}\;.
\]
Since any two bases differ by a $\GL(n,\bbR)$ transformation, $\coframe(M)$ 
is a principal $\GL(n,\bbR)$ bundle, \emph{i.e.}, the transition functions are 
$\GL(n,\bbR)$ valued.

The best known sub-bundle of $\coframe(M)$ is the bundle of
orthonormal frames $\Orthframe(M)$.  This bundle may be defined via the use
of the metric $ds^2$ by defining it to be
\[
  \Orth(M) =
  \left\{ \omega \in \coframe(M)\;|\; ds^2 = 
  \omega^i\otimes\omega^i\right\} \subset \coframe(M)\;.
\]
From the definition it is clear that any two coframes at a point $x\in
M$ differ by an orthogonal transformation therefore the bundle of
orthonormal coframes is a principal bundle with structure group
$\Orth(n)$.

A $G$-structure is a reduction of the coframe bundle to a principal 
bundle with structure group $G$. For our our purposes it is convenient to 
give a local description of a $G$-structure. Assume that we are given an 
open cover of $M$ by open sets $\{V_\alpha\}$ and a collection of 
coframes $\omega_\alpha$ defined on $V_\alpha$. We assume that on a 
non-empty overlap $V_\alpha \cap V_\beta$ one has
\[
    \omega_\alpha = \gamma_{\alpha\beta} \omega_\beta
\]
where the transition functions are $G$-valued
$\gamma_{\alpha\beta}: V_\alpha \cap V_\beta \to G$. We require the 
transition functions to satisfy the usual cocycle conditions.
Given the transition functions $\{\gamma_{\alpha\beta}\}$ one constructs a 
principal fiber bundle by locally patching the sets $V_\alpha \times G$ 
using the transition function. For $x\in V_\alpha \cap V_\beta$,
if $(x,g_\alpha) \in V_\alpha \times G$ 
and $(x,g_\beta)\in V_\beta \times G$ then we identify $(x,g_\alpha)$ and 
$(x,g_\beta)$  if
\[
    g_\beta = g_\alpha \gamma_{\alpha\beta}(x) \;.
\]
The principal bundle thus constructed is called a {\em
$G$-structure\/}.

$G$-structures have a globally defined canonical
$1$-form\footnote{This is sometimes called the soldering form.} that
distinguishes a $G$-structure from a generic principal $G$-bundle.  We
observe that $g_\alpha \omega_\alpha = g_\beta \omega_\beta$ therefore
we have a $n$ globally defined forms that we can put into a
column vector and call them $\omega$.  When restricted to $V_\alpha
\times G$, the forms $\omega$ may be written as
\[
   \omega|_{V_\alpha \times G} = g_\alpha \omega_\alpha\,.
\]
Finally we note that there exists a local section
$s_{\alpha}:V_{\alpha}\to \coframe(M)$ such that $s^{*}_{\alpha}\omega
= \omega_{\alpha}$.  If $\pi:\coframe{M} \to M$ is the projection
defining the bundle then it is not true that $\omega =
\pi^{*}\omega_{\alpha}$.

There are a variety of notable $G$-structures:
\begin{itemize}
	\item $G=\Orth(n)$ gives Riemannian structures and is equivalent
	to specifying a riemannian metric.  \item $G=\SL(n)$ is
	equivalent to prescribing a volume element, \emph{i.e.}, an
	orientation. 
	
	\item $G=\SOrth(n)$ gives orientable Riemannian structures. 
	\item $G=\{e\}$, the trivial group, is equivalent to
	specifying a global coframe, \emph{i.e.}, the manifold is
	parallelizable.  These are called $\{e\}$-structures.
\end{itemize}
On even dimensional manifolds, $\dim M = 2m$, we have:
\begin{itemize}
	\item $G=\Spx(2m,\bbR)\subset \GL(2m,\bbR)$ gives almost symplectic
	structures.

	\item $G=\GL(m,\mathbb{C}) \subset \GL(2m,\bbR)$ gives almost
	complex structures.

	\item  $G=\U(m)=\SOrth(2m)\cap\Spx(2m,\bbR)\subset \GL(2m,\bbR)$ 
	gives  almost hermitian structures.
\end{itemize}

\section{Riemannian Geometry of Lie Groups}
\label{sec:Liegroups}

Assume $G$ is a connected real compact Lie group of dimension
$n$.  Choose an orthonormal coframe of Maurer-Cartan forms
$\omega^{i}$ for an $\Ad(G)$-invariant metric.  The Maurer-Cartan
equations are
\begin{equation}
    d\omega^{i} = - \half f^{i}{}_{jk}\omega^{j} \wedge\omega^{k}\,,
    \label{eq:MC}
\end{equation}
where $f_{ijk}$ are the totally skew symmetric structure constants for 
the Lie algebra $\lieg$ of $G$. The invariance of the metric tells us 
that the adjoint group $\Ad(G)$ is a subset of $\SOrth(n)$.
Comparing with (\ref{eq:cartan1}), using the skewness of $f_{ijk}$ and 
using the uniqueness of the riemannian connection we 
immediately conclude that
\begin{equation}
    \omega_{ij} = -\half f_{ijk}\omega^{k}\,.
    \label{eq:Gconnection}
\end{equation}
Using (\ref{eq:cartan2}) we see that the riemannian curvature of the Lie 
group is given by
\begin{equation}
    R_{ijkl} = \half f_{ijm} f_{mkl} 
    + \frac{1}{4}\left( f_{imk}f_{mjl} - f_{iml}f_{mjk}\right)
    = \frac{1}{4}f_{ijm} f_{mkl}\,
    \label{eq:Gcurv}
\end{equation}
where the last equality was obtained by using the Jacobi identity. 
You should compare the structure of the second term above with
(\ref{eq:RRtil}).  Finally we observe that $f_{ijk}$ is covariantly
constant with respect to the riemannian connection because of the
Jacobi identity:
\begin{eqnarray}
    \nabla f_{ijk} & = & -\omega_{im}f_{mjk} - \omega_{jm}f_{imk}
    - \omega_{km}f_{ijm}\,,
    \nonumber  \\
     & = & +\half \left(
     f_{iml}f_{mjk} + f_{jml}f_{imk}
    + f_{kml}f_{ijm} \right) \omega^{l}\,,
    \nonumber \\
    & = & 0\,.
    \label{eq:covf}
\end{eqnarray}

All the equations above are on $G$.  The corresponding
expressions for the connection and the curvature on the coframe bundle
$\coframe(G) = G \times \SOrth(n)$ are different.

\section{Torsion}
\label{sec:torsion}

We work in the bundle of orthonormal frames $\Orthframe(M)$ on the
manifold $M$ with riemannian connection $\omega_{ij}$ that satisfies
the Cartan structural equations (\ref{eq:cartan1}) and
(\ref{eq:cartan2}).  We have the option of considering a second
orthogonal connection $\phi_{ij} = \omega_{ij} + C_{ijk}\omega^{k}$
where $C_{ijk}=-C_{jik}$.  With respect to this new metric compatible
connection on $\Orthframe(M)$, the Cartan structural equations are
\begin{eqnarray}
    d\omega^{i} & = & -\phi_{ij}\wedge \omega^{j} + \half T^{\phi}_{ijk} 
    \omega^{j}\wedge \omega^{k}\,,
    \label{eq:cartan1tor}  \\
    d\phi_{ij} & = & -\phi_{ik}\wedge \phi_{kj}
    + \half R^{\phi}_{ijkl}\omega^{k}\wedge\omega^{l}\,.
    \label{eq:cartan2tor}
\end{eqnarray}
In the above the torsion $T^{\phi}_{ijk}$ is related to the ``contorsion'' 
$C_{ijk}$ by
\begin{equation}
    T^{\phi}_{ijk} = - (C_{ijk}-C_{ikj})\,.
    \label{eq:contorsion}
\end{equation}
The curvatures for the two connections are related by
\begin{equation}
    R^{\phi}_{ijkl} = R^{\omega}_{ijkl}
    +(\nabla^{\omega}_{k}C_{ijl} - \nabla^{\omega}_{l}C_{ijk})
    +(C_{imk}C_{mjl} - C_{iml}C_{mjk})\,.
    \label{eq:phicurv}
\end{equation}
It is also possible to express the above in terms of the covariant
derivative $\nabla^{\phi}$ with respect to the connection $\phi$.  To
simplify matters we express the above \emph{only} for the case where
$C_{ijk}$ is totally antisymmetric:
\begin{equation}
    R^{\phi}_{ijkl} 
    -(\nabla^{\phi}_{k}C_{ijl} - \nabla^{\phi}_{l}C_{ijk})
    = R^{\omega}_{ijkl}
    - 2 C_{ijm}C_{mkl}
    -(C_{imk}C_{mjl} - C_{iml}C_{mjk})\,.
    \label{eq:phicurv1}
\end{equation}

\providecommand{\href}[2]{#2}\begingroup\raggedright\endgroup

\end{document}